\documentclass[a4paper,12pt]{article}
\usepackage{amsfonts}
\usepackage{amssymb}
\usepackage{makeidx}
\usepackage[english]{babel}
\usepackage[dvips]{graphicx}
\newcommand{\nn}{ \noindent }

\newcommand{\fine}{ $ \Box  $}

\long\def\symbolfootnote[#1]#2{\begingroup%
\def\thefootnote{\fnsymbol{footnote}}\footnote[#1]{#2}\endgroup}

\makeindex

\begin{document}

\newtheorem{defn}{Definition}
\newtheorem{theo}{Theorem}
\newtheorem{ex}{Example}
\newtheorem{prop}[theo]{Proposition}
\newtheorem{lemma}[theo]{Lemma}
\newtheorem{coroll}[theo]{Corollary}
\newtheorem{conj}[theo]{Conjecture}
\newtheorem{note}[defn]{Point}
\newtheorem{apprx}[defn]{\underline{Approximation}}

\newtheorem{tbd}{--> To be done}


\title{Optimal equilibria of the best shot game\thanks{P. P. acknowledges support from the project Prin 2007TKLTSR "Computational markets design and agent--based models of trading behavior".}}
\author{Luca Dall'Asta\thanks{%
The Abdus Salam International Centre for Theoretical
Physics, Strada Costiera 11, 34014 Trieste, Italy}
\and
Paolo Pin\thanks{%
Dipartimento di Economia Politica, Universit\'a degli Studi di Siena, Piazza San Francesco 7, 53100 Siena, Italy}
\and
Abolfazl Ramezanpour\thanks{%
Dipartimento di Fisica, Politecnico di Torino, Corso Duca degli Abruzzi 24, 10129 Torino, Italy}}
\maketitle

\begin{abstract}
We consider any network environment in which the ``best shot game'' is played.
This is the case where the possible actions are only two for every node ($0$ and $1$),
and the best response for a node is $1$ if and only if all her neighbors play $0$.
A natural application of the model is one in which the action $1$ is the purchase of a good, which is locally a public good, in the sense that it will be available also to neighbors.
This game typically exhibits a great multiplicity of equilibria.
Imagine a social planner whose scope is to find an optimal equilibrium, i.e. one in which the number of nodes playing $1$ is minimal.
To find such an equilibrium is a very hard task for any non--trivial network architecture.
We propose an implementable mechanism that, in the limit of infinite time, reaches an optimal equilibrium, even if this equilibrium and even the network structure is unknown to  the social planner.
\end{abstract}

\nn {\bf JEL classification code:} C61, C63, D85, H41.

\bigskip

\nn{\bf Keywords:} networks, best shot game, simulated annealing.

\newpage

\section{Introduction.}

Take an exogenous network in which otherwise homogeneous players (nodes) play a public good game, which is the one defined \emph{Best shot game} in Galeotti et al. (2010).\footnote{Galeotti et al. (2010) give this name in Example 2 and use it throughout the paper.
The name \emph{Best shot game} comes from Hirschleifer (1983), where it is however described as a non--network game.}
The best shot game is a discrete case, with restricted strategy profiles and satiated utilities, of the model in Bramoull\'e and Kranton (2007) and of the second stage of the game in Galeotti and  Goyal (2008).
The action of each node $i$ is an effort $x_i$ and her payoff depends on the aggregate effort of herself and that of her neighbors, minus some cost for her own effort.

Here we restrict strategy profiles to the two specialized actions: $x_i \in \{ 0,1\}$.\footnote{%
One result in Bramoull\'e and Kranton (2007) is actually that, even when the possible actions of nodes are continuous, in a \emph{stable} equilibrium every agent would play either $0$ or a fixed value $e^*>0$ which can be normalized to $1$.}
In this way $\vec{x}$, a vector of specialized actions whose length is given by the number of nodes, will characterize any possible configuration of the system.
We will consider the class of incentives such that, in Nash equilibrium (NE), agent $i$ will play action $x_i$ according to the following rule:
\begin{equation}
\left\{
\begin{array}{ll}
x_i=1 & \mbox{if $x_j=0$ for all neighbors $j$ of node $i$;}  \\
x_i=0 & \mbox{otherwise.}
\end{array}
\right.
\label{rule}
\end{equation}
We will study all the NE of the game: that is all those action profiles in which, for any link, not both nodes of the link put in effort $1$;
but at the same time for any node, if we consider the set including itself and its neighborhood, at least one node in this set puts in effort $1$.
Mathematically, the subset of nodes playing $1$ in a NE will then be a \emph{maximal independent set} of the network, as it is called in graph theory.

The next example will give some insight on the maximal independent sets, our NE, for simple networks.

\begin{ex} A network of $9$ nodes. \label{Ex1} \end{ex}

\begin{figure}[h]
\begin{center}
\includegraphics*[width=10cm]{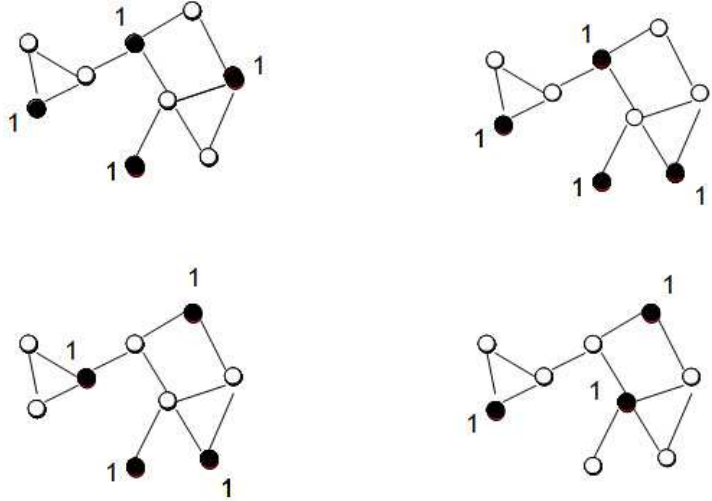}
\caption{Four NE for a $9$--nodes network.}
\label{9nodes}
\end{center}
\end{figure}

Figure \ref{9nodes} shows four possible NE for the same network of $9$ nodes.
Black nodes are those playing $1$, while all the others are playing $0$.
The bottom--right NE is the only one in which only three nodes play action 1.
If we assume action $1$ to be a costly action, interpreting it as the purchase of a local public good, then the bottom--right NE is socially optimal, at least regarding costs. \fine

\bigskip

By considering this last example, a first intuition is that when more connected nodes play $1$, then the number of $1$--players in equilibrium is reduced.
The extremal case of this will happen on a star--shaped network, as shown in the next example.

\begin{figure}[h]
\begin{center}
\includegraphics*[width=8cm]{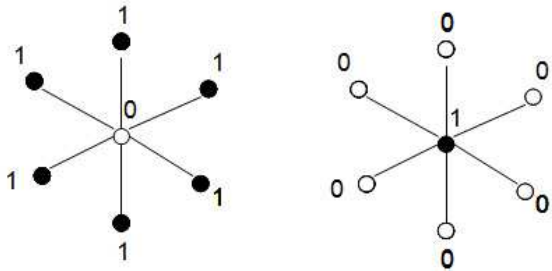}
\caption{The two NE of a star network.}
\label{star_fig}
\end{center}
\end{figure}

\begin{ex} The star. \label{star} \end{ex}
It is easy to see that the star has only two maximal independent sets (see Figure \ref{star_fig}):
one in which the center alone plays $1$, and another one in which the spokes do so.
If we are looking for efficiency (defined as fewer $1$s, which are supposed to be costly) it is very easy to find that the first case is the best one.
Suppose that we are in the \emph{bad} NE (spokes exerting the costly effort), then a social planner could shift to the \emph{good} equilibrium by incentivating a contribution from the center.
When the center is contributing, then, by best response, all spokes stop doing so.
This mechanism will be formalized in the next section, but the idea is that of incentivating a contribution from agents that were not contributing in a NE, thus the system will move to a new NE, which may reduce the social cost of being in equilibrium. \fine

\bigskip


The problem of finding all maximal independent sets of a general network is however not an easy one and will be discussed in Section \ref{speed}.
This problem is actually NP--hard,\footnote{%
An optimization problem is NP-hard if it is as difficult as any problem in the NP--complete (\emph{non--deterministic polynomial}) class.
Consider a general problem whose object (input) is characterized by a certain size $N$ (as could be the number of nodes in our case).
Here is given a non--rigorous definition:
The problem is called NP--complete if there is no algorithm that can find a solution to the problem, for any possible input of size $N$, in a time that grows at most polynomially in $N$.
An NP--complete problem is one in which the time required to find a solution typically grows exponentially in $N$.
In practice this means that, even if a good computer can solve the problem in a reasonable time for $N=1.000$, the case $N=10.000$ may take years to be solved.}
as is 
the problem of finding those maximal independent sets with more or less nodes playing $1$.
In a companion paper, Dall'Asta, Pin and Ramezanpour (2009),
we discuss these aspects in more detail for a particular class of random networks.
The next example may give a hint of this.

\begin{ex} A regular random network. \label{reg_rand_ex} \end{ex}

\begin{figure}[h]
\begin{center}
\includegraphics*[width=6cm]{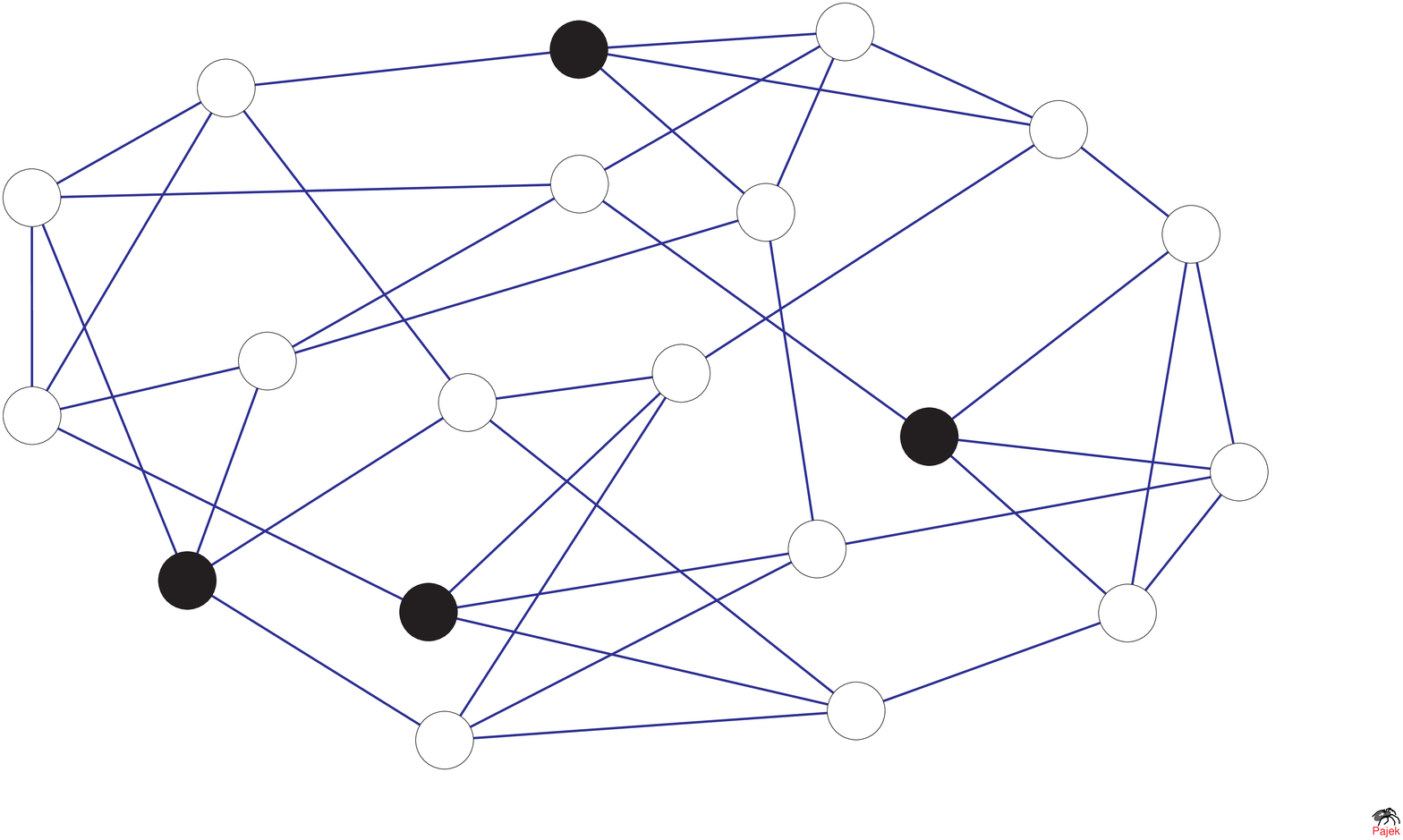}
\includegraphics*[width=6cm]{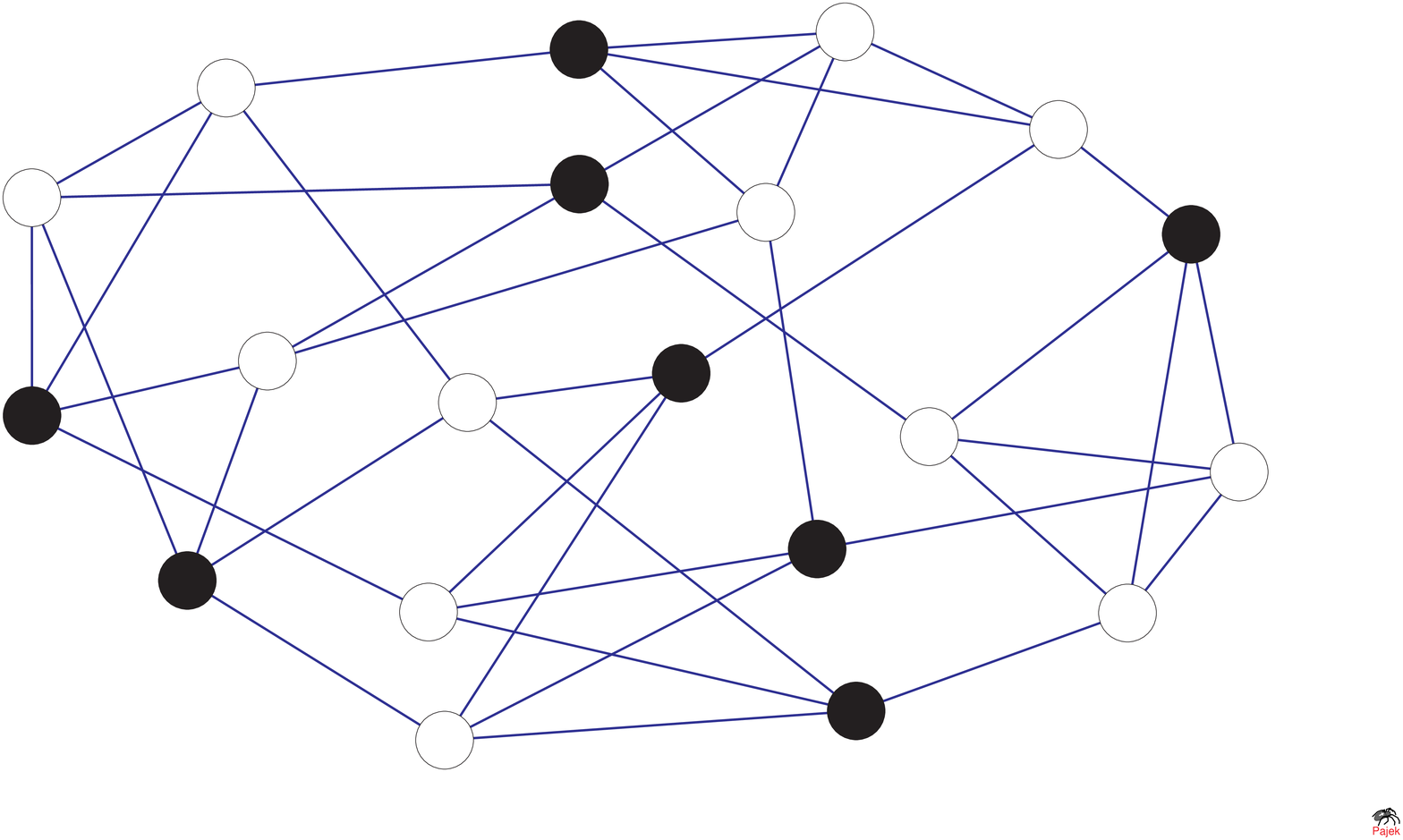}
\caption{Two NE for the same regular random network of $20$ nodes and degree $4$.
Picture is obtained by means of the software Pajek (http://pajek.imfm.si/).}   \label{reg_rand20}
\end{center}
\end{figure}

Consider the regular random network illustrated (twice) in Figure \ref{reg_rand20}.
It has $20$ nodes, and each of them has exactly $4$ links.
In this case we cannot propose any strategy that targets as contributors those nodes with many links, as could be suggested from previous examples.
This network in particular has $128$ equilibria: $2$ (one is in Figure \ref{reg_rand20}, left) with $4$ nodes contributing, $25$ with $5$, $58$ with $6$, $42$ with $7$, and only $1$ (Figure \ref{reg_rand20}, right) with $8$ nodes contributing. \
In Dall'Asta, Pin and Ramezanpour (2009) we consider such networks consisting of a large number of nodes, and we use an analytic
approach to compute the approximate number of NE as a function of the fraction of contributors.\footnote{%
By adopting a mean field analysis, L\'opez--Pintado (2008) identifies the mean fraction of contributors for a typical NE.}
The predictions are very accurate when the number of nodes is large, but search algorithms are unable to successfully explore in finite time the large deviations predicted by the theory (this problem is also NP--hard). \
For small networks, even if regular random, there is a lot of variability.
Other networks of $20$ nodes and degree $4$, generated with the same random process, have completely different distributions.
The only way to find all the equilibria in a particular network is to control all the $2^{20} \simeq 10^6$ possible pure strategy profiles. \fine

\bigskip

From the point of view of economics, the rule specified in (\ref{rule}) is not  \emph{behavioral} and could be justified by several modelling choices with rational agents.
Up to now we have defined (pure) Nash equilibria without explicitly defining actions and payoffs; this however could easily be done.
One possibility is the following.
Any agent attributes utility $v$ to a homogeneous good, if she has access to it (independently of whether it is provided by herself or by any of her neighbors), and her utility is satiated by one unit of it.
Finally, the cost of providing the good is a positive value $c<v$.
Since utilities are satiated, and in equilibrium every agent has local access to the good, then considering efficiency from the point of view of minimal aggregated costs is enough to achieve global efficiency.
In our model agents consider only local spillovers and exclude any externality from any other non-neighbor player.
In this sense the network structure formalizes the range of the externalities.
Note however that, because of satiation, the utility of agents is not linear in the contribution effort of neighbors, so that our model is not included in the class of games analyzed by Ballester et al. (2006), hence it cannot be solved with the help of Bonacich centrality.
Bramoull\'e and Kranton (2007) consider non--satiated utility functions and find the typical public--good discrepancy between efficient strategy profiles and equilibria.
A general class of games that includes the one from Bramoull\'e and Kranton (2007) is analyzed in Bramoull\'e et al. (2009), however also their class does not include our non--satiated utilities.

In next section we will define formally the general \emph{best reply} mechanism that we consider, and that we implement also in the numerical simulations.
From a theoretical point of view, it may seem that we exclude full rationality when we assume that agents respond to changes with a best response rule that considers only the present configuration but is myopic and not strategic on possible future new changes.
Consider, however, that another explanation for  agents not being interested in future expected payoffs is a high rate $\delta$ of temporal discount.

The kind of situation we have in mind is that of every agent deciding whether or not to exert a fixed costly effort that is beneficial to herself and also to her neighbors, so that a typical situation of free riding incentives arises.
This could be the case with farmers or firms adopting new technologies, with an information network and a cost for possible failures.\footnote{%
This is the application proposed in Bramoull\'e and Kranton (2007), where they cite the applied model in Foster and Rosenzweig (1995).}
Another application could be that of several municipalities in a given region; the public good could be a library or a fire brigade, and two municipalities are linked if the public good in one of them makes the same public good undesirable in the other one because of geographical proximity.
Finally, since the mechanism we propose requires low costs of shifting between strategies and repeated interaction, a good application could be that of a big firm encouraging people to share cars in order to minimize parking places.
Action $1$ would mean `take the car' and an employee would play $0$ if a friend gives her a lift.
Generally, in any of these applications there could be a planner whose objective could reasonably be that of minimizing costs.

Suppose that the planner considers all possible NE of the game (all maximal independent sets of the network) and wants to minimize among them the number of nodes exerting effort $1$ (i.e. find a maximal independent set of minimal cardinality: MNE).
She could impose the proper action on the agents, and the resulting configuration, being a NE, would be stable without imposing more incentives.
Suppose, however, that the planner does not know such an optimal distribution (remember that the theoretical problem is typically a complex one) or that moreover she may not even know anything about the network.
Assuming that we also have a time dimension, our question is: would it still be possible for the planner to build a \emph{mechanism}
that would incentivate the agents to move towards an optimal MNE?\footnote{%
We will use the term \emph{mechanism} to differentiate it from \emph{algorithm}.
While the latter is intended as a computational technique, the former is a plausible implementation of any single step of such a technique into a real system, also allowing the interaction of self--interested agents.}
Our answer is only theoretical but positive: at the limit of infinite time such a mechanism exists, and it will lead to a MNE with probability $1$.

What we assume is that the social planner's goal is to minimize the costs of a NE, when she has the possibility of incentivating players' actions out of equilibrium, but she is not able to modify the structure of the network.
It is clear that if the planner had the possibility of changing the network structure, directly or by incentives, at a reasonable cost (as is the case considered on a different network game by Haag ad Lagunoff (2006)) then the problem would look very different.
It would be enough to approximate a star--like configuration such as the one analyzed in Example \ref{star}, and the solution would easily be found.

In the next section we show how we obtain our result.
We show that our setup is included in the hypothesis of a theorem first proved in Geman and Geman (1984) and presented here in Appendix \ref{Geman}.
The proof of this equivalence is based on three lemmas, whose proofs are in Appendix \ref{Proofs}.
Section \ref{speed} analyzes, mainly by means of numerical simulations, how the simulated annealing approach that we propose performs in two very different network structures: regular random networks and scale free networks.
We conclude the paper with Section \ref{conclusion}.


\section{Main result}

The mechanism we study is defined in discrete time ($t=1,2,3,\dots$).
At every time step the configuration $\vec{x}_t$ of nodes' actions satisfy condition (\ref{rule}) for every node, and hence is a NE.
Suppose then that at time $1$ the system is in a NE, so that $x_{i,1} \in \{ 0 ,1\}$ is a best response for every agent $i$, as specified in (\ref{rule}).
The planner does not know anything about the network, the only thing she observes at any step $t$ in time is the action of each player and hence the aggregate number $M_t = \sum_{i} x_{i,t}$ of agents playing $1$.
At every time step, she picks an agent $i_t$ playing $0$, at random with uniform probabilities, and induces her to flip her strategy to $1$.\footnote{%
This can easily be done through incentives.
The reason why the planner is looking for a minimum could be that she is financing all the agents exerting effort; in this case she could raise her contribution to the agent up to the desired threshold level.}
Let us call this transition $F$.
The transition $F$ is defined only from a NE $\vec{x}$ to a non--NE $\vec{x'}$.
It defines a Markov chain across all $\{0,1\}$ vectors $\vec{x}$. 
In consequence of this flip, all the other nodes in the network will change their strategy according to the \emph{best response} rule defined here below.

Consider the subset of unsatisfied agents in a non--NE configuration, i.e. any agent for which condition (\ref{rule}) is violated, either because she plays $0$ and also all her neighbors do, or because she plays $1$ and at least one of her neighbors do the same.
If we apply transition $F$ to a node $i_t$ who was originally playing $0$, then the set of unsatisfied nodes includes always elements different from $i_t$; as in a NE there is always at least one node $j$ playing $1$ around any node $i_t$ playing $0$.
The basic step of the best response rule, is iterated by picking with uniform probabilities one of the unsatisfied nodes, different from $i_t$, and flipping her strategy.
Let us call this transition step $B$.
This basic step $B$ clearly defines a Markov chain across all $\{0,1\}$ vectors $\vec{x}$, whose absorbing states are NE.
In Proposition \ref{main} we show that if we start from a NE, we apply $F$ once, and then we iterate $B$, we reach with probability $1$, and with a limited number of steps, a new $NE$.
We show also that, for the scope of this result, we can discard without loss of generality the possibility of synchronous updating.
It is clear that, in the assumptions of the model, $F$ is induced by the planner, while the iteration of $B$ is obtained from the endogenous adaptation of the agents, as long as they are not all satisfied.

When the system is stable again, i.e. again in a new NE, the planner will observe a new configuration $\vec{x}^{new}_t$ and the new aggregate quantity of $1$'s, call it $M^{new}_t$.
The planner will accept the new configuration with probability
\begin{equation}
\left\{
\begin{array}{ll}
1 & \mbox{if } M^{new}_t<M_t \ \ ; \\
t^{- \epsilon \left( M^{new}_t-M_t \right) } & \mbox{otherwise,}
\end{array}
\right.
\label{montecarlo}
\end{equation}
where $\epsilon >0$ is a constant. The second probability in (\ref{montecarlo}) identifies the level of rejection of non--improving changes.

We start by proving that $\vec{x}^{new}_t$ is always a NE for any $t$ (see Lemma \ref{allNE} below).
If the planner accepts the new configuration, then $\vec{x}_{t+1} = \vec{x}^{new}_t$ and $M_{t+1} = M^{new}_t$, otherwise she will impose reverse incentives so that we return to the original configuration,\footnote{%
This can be done by reverting all incentives to the nodes who changed; they are, by following Lemma \ref{localNE}, restricted to a local neighborhood.}
i.e. $\vec{x}_{t+1} = \vec{x}_t$ and $M_{t+1} = M_t$.

In the limit $t \rightarrow \infty$, the second probability in (\ref{montecarlo}) goes to $0$ and the mechanism will converge to any member of a precise subset of NE.
Call the subset of such possible NE \emph{local minima}.\footnote{%
It is also possible that the mechanism, at the limit $t \rightarrow \infty$, alternates between more than one single NE, if all of them have the same number of $1$'s.
Without loss of generality, such subsets of NE can simply be included among \emph{local minima}.}
Every MNE is also a local minimum.
The question is whether the local minimum in which the process ends is also a MNE.
The aim of this paper is to show under which conditions the answer is positive.

The structure of the proof is the following.
We show that we meet the conditions required for the application of a known theorem.

\begin{lemma} \label{allNE} If we start from a NE and invert the action of one node from $0$ to $1$, then the best response rule of all the other nodes in the network will imply a new NE.
\end{lemma}

\begin{lemma} \label{localNE} If we start from a NE and invert the action of one node from $0$ to $1$, then the best response rule of all the other nodes in the network will be limited to the neighborhood of order $2$ of the original node (i.e. the change is only local).
\end{lemma}

\begin{lemma} \label{anyNE} It is possible to reach any NE from any other NE with a finite number of the following procedures: flip the action of a single node from $0$ to $1$ (transition $F$) and obtain, by iterated best response of the nodes (transition step $B$), a new NE.
\end{lemma}

\begin{prop} \label{main}
The probability $\pi (\epsilon)$ that the mechanism ends in a MNE, in the limit $t \rightarrow \infty$, is strictly positive for any $\epsilon>0$; it is decreasing in $\epsilon$; and finally, there exists an $\bar{\epsilon}>0$ such that, for any $\epsilon<\bar{\epsilon}$, we have that $\pi (\epsilon) = 1$ independently on the initial conditions.
\end{prop}

{\bf Proof:}
consider the set $\Omega$ of NE of a given finite network, which is a subset of all the $\{0,1\}$ vectors $\vec{x}$, and call $N_{\Omega} \equiv |\Omega|$ its finite cardinality.
Call $|\vec{x}|$ the number of agents playing $1$ in an equilibrium $\vec{x} \in \Omega$, and define $U^* \equiv \max \{ |\vec{x}| : \vec{x} \in \Omega \}$, $U_* \equiv \min \{ |\vec{x}| : \vec{x} \in \Omega \}$,  and $\Delta \equiv U^* - U_*$. \\
If we apply first $F$ to any $\vec{x} \in \Omega$ and then we iterate $B$, then by the proof of Lemmas \ref{allNE} and \ref{localNE}, in a finite number of iterations we reach a new NE $\vec{x'} \in \Omega$, with $\vec{x'} \ne \vec{x}$
This defines a stochastic process $X$ between the states of $\Omega$ which is ergodic because of Lemma \ref{anyNE}.

Then, we are in the conditions of Theorem B in Geman and Geman (1984) (see Appendix \ref{Geman}), and $\bar{\epsilon} \equiv \frac{1}{N_{\Omega} \Delta}$. \fine

\bigskip

The lemmas are proven in Appendix \ref{Proofs}, by applying the discrete mathematics of network theory.
Lemmas \ref{allNE} and \ref{localNE} 
also guarantee that the proposed mechanism is well defined.

The main proposition is obtained by including our setup in the general hypothesis of the theory of \emph{simulated annealing}, first proposed and formalized in Kirkpatrick, Gelatt and Vecchi (1983).
Simulated annealing is a heuristic algorithm based essentially on the increasing rejection probability in a Monte Carlo step, as the probability $t^{- \epsilon \left( M^{new}_t-M_t \right) }$ in (\ref{montecarlo}), for our case.
Simulated annealing works exactly as described above, finding a global minimum of a certain function, avoiding local minima.
Theory tells us that, if the number of possible configurations is finite, and it is possible to reach any configuration from any other with basic steps, then a generalization of the above proposition holds.
The rigorous proof that applies to our model can be found in Theorem B of Geman and Geman (1984), which we discuss in Appendix \ref{Geman}.
The original proof takes various pages, its intuition is that we are analyzing a Markov chain of finite possible configurations (all the NE of the game) which is ergodic for any finite $t$.

In our case, we consider all the NE as the possible states of the system; they are finite because the network is finite.
Lemmas \ref{allNE} and \ref{localNE} define a stochastic process between the states of the system, and this process is ergodic by Lemma \ref{anyNE}.
We thus meet the conditions that apply in Appendix \ref{Geman}.

\section{Accuracy vs. speed of convergence} \label{speed}

The mechanism that we propose reaches an optimal outcome with probability $1$ but is extremely time consuming.
In this section we discuss how in some cases the choice of a faster mechanism (i.e. a higher $\epsilon$) could be useful if we are looking for almost optimal solutions in shorter time.
However, the trade--off between accuracy and speed of convergence is very hard to compute in general.
Simple adaptations of the mechanism may not be useful at all in some case, as we show here below by means of computer simulations.

We run simulations on random regular networks, as the one in example \ref{reg_rand_ex}, and on scale--free networks, as the one illustrated in the following example.\footnote{%
It is well known that random regular and scale--free networks do not have some of the properties, as \emph{clustering} or \emph{assortativity}, that real world networks have (see Newman (2003) and Jackson and Rogers (2007) for more discussion), and that other models would be more realistic in generating large random networks.
However, as we are working with small networks of $20$ nodes,  the two models that we are using provide the necessary distinction between a homogeneous and a heterogeneous distribution of links, and differentiations on other dimensions are irrelevant.}

\begin{ex} A scale--free network. \label{scale_free_ex} \end{ex}

\begin{figure}[h]
\begin{center}
\includegraphics*[width=6cm]{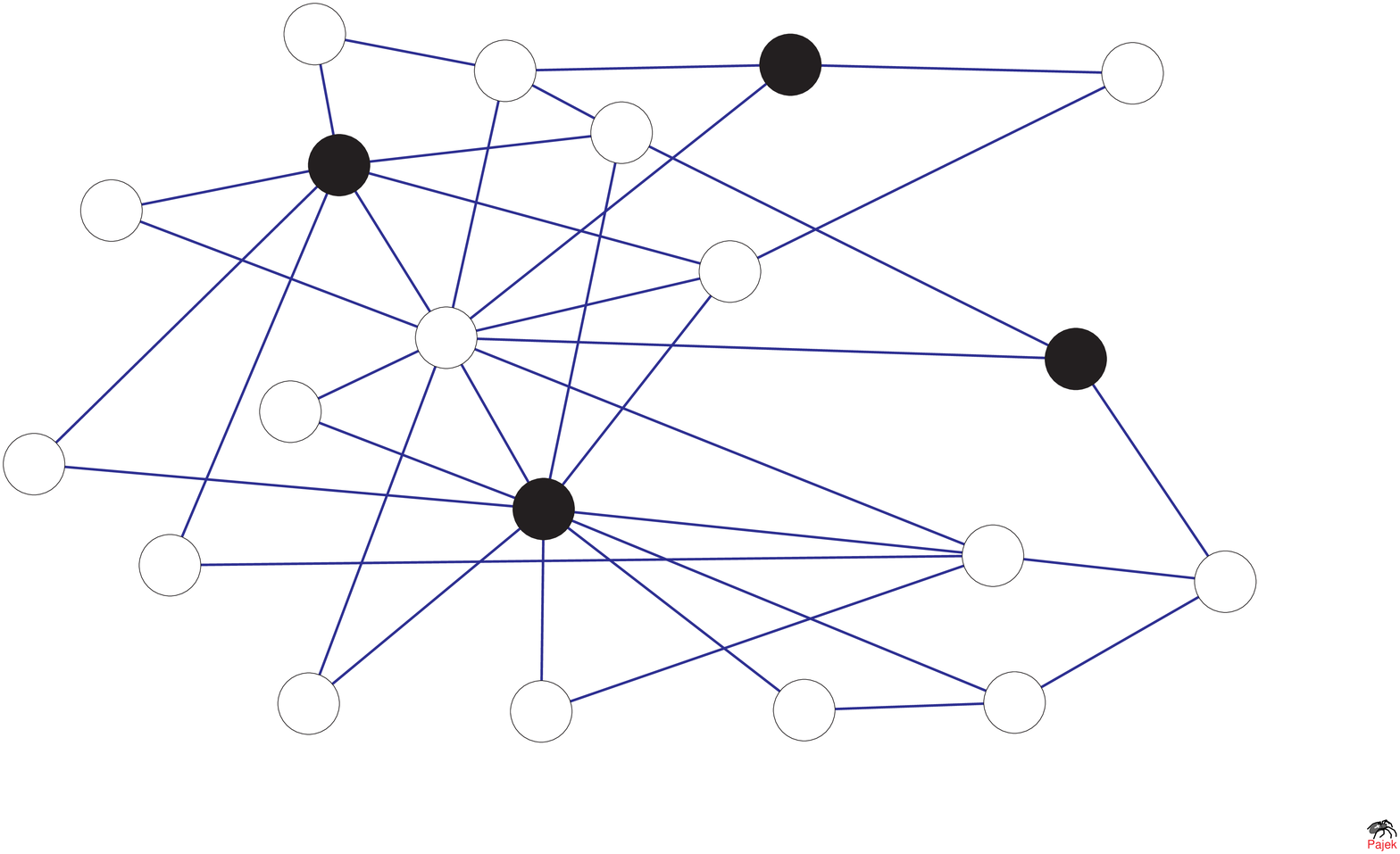}
\includegraphics*[width=6cm]{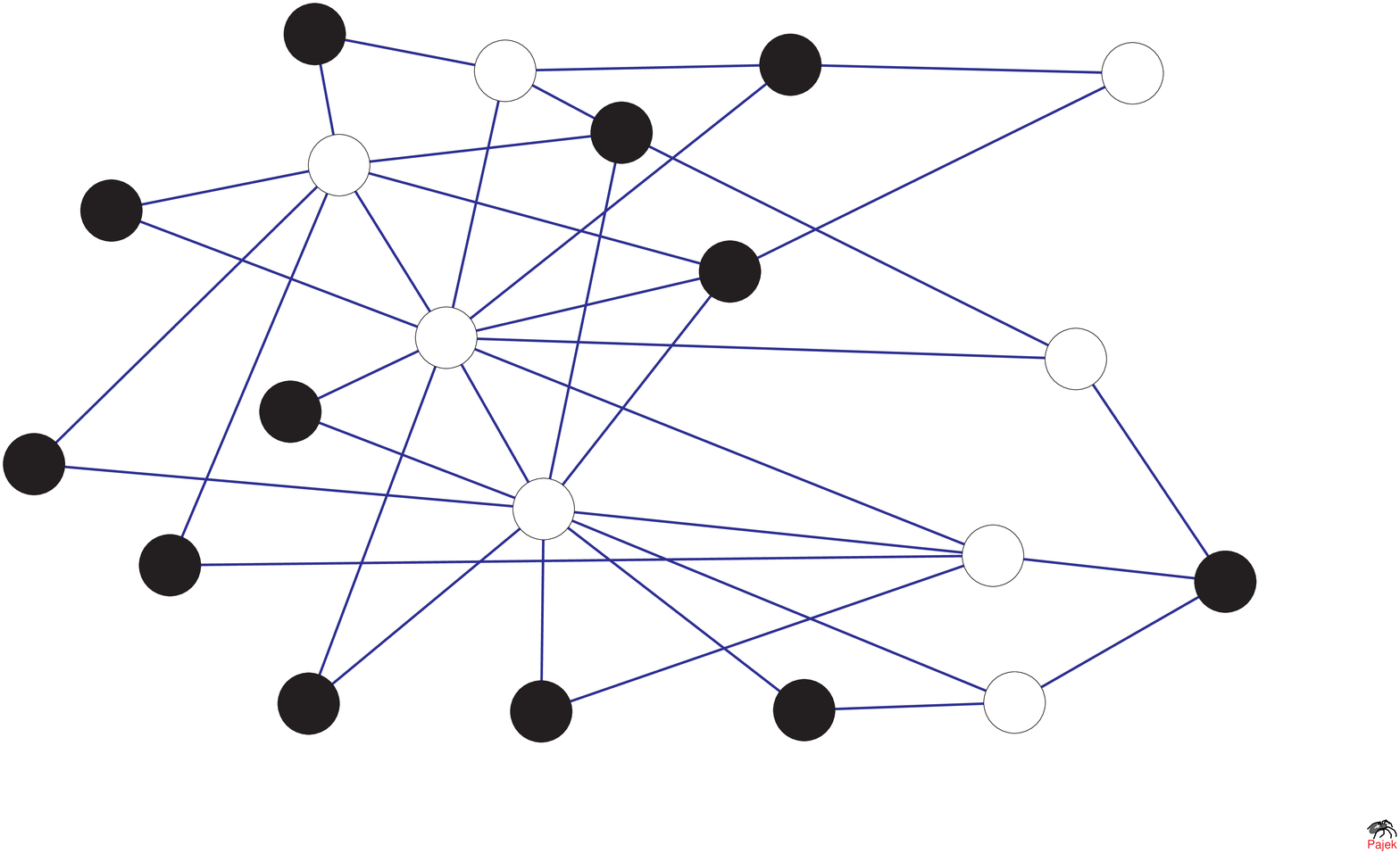}
\caption{Two NE for the same scale--free network of $20$ nodes and average degree $4$.
Picture is obtained by means of the software Pajek (http://pajek.imfm.si/).}   \label{scale_free20}
\end{center}
\end{figure}

Consider the random scale--free network illustrated (twice) in Figure \ref{scale_free20}.
It has been generated with the simple algorithm proposed in Albert and Barabasi (1999):
it has $20$ nodes, and they have an average degree of exactly $4$ links. \
This network in particular has $48$ equilibria: $2$ (one is in Figure \ref{scale_free20}, left) with $4$ nodes contributing, $2$ with $5$, $6$ with $6$, $5$ with $7$, $6$ with $8$, $9$ with $9$, $13$ with $10$, $4$ with $11$, and only $1$ (Figure \ref{scale_free20}, right) with $12$ nodes contributing.
Other such networks of $20$ nodes and average degree $4$, generated with the same algorithm, have completely different distributions. \
As there is heterogeneity in the distribution of links, a good strategy to find efficient equilibria could be that of targeting as contributors those nodes with many links.
However, in this way we may not find the \emph{good} equilibrium with only $4$ contributors, where a node with $10$ links is not contributing, while two of the four contributors have only $3$ links. \
Also in this class of random networks we are able to find and compare all the equilibria only by controlling all the $2^{20} \simeq 10^6$ possible pure strategy profiles. \fine

In the simulations we do the following.
First we generate a random network with one of the two models;
then we count all the $N_{\Omega}$ equilibria of that network out of all the pure strategy profiles,
and from this information we can easily obtain also $\Delta \equiv U^* - U_*$, for that particular network.
Then we compute the Markov matrix induced, on the set ${\Omega}$ of NE, by the application of $F$ and the iteration of $B$ to any element of that set.
From this we get the information about which ones of the NE of that network are also local minima of the Markov process,
and also about which ones are local minima but not MNE.
Finally, we run simulated annealing on that matrix with different values of $\epsilon>\bar{\epsilon}$.

In principle, as we obtain the Markov matrix, we could apply theoretical results (see e.g. the lectures of Catoni (1999)) to approximate the accuracy and the speed of convergence that any $\epsilon$ would give.
The problem is that any single different network, obtained with the same model, may have a completely different number and distribution of the NE.
The theoretical results could be applied only for a particular network or for very specific and completely symmetric classes of networks, for which the problem of finding the MNE is however a trivial one to solve.

We run the simulation described above with $50$ random regular networks of $20$ nodes and degree $4$, and with $50$ scale--free networks of $20$ nodes and average degree $4$. We use a log--grid of $\epsilon$'s that are multiple of $\bar{\epsilon}$.
The factor of multiplicity ranges from $10$ to $1000$.
For any run of the simulated annealing we report the time needed to find a NE.\footnote{%
We assume that the simulated annealing algorithm that we run \emph{converges} when it does not change for $10^4$ steps, and a threshold is set at $10^7$ steps.}
We report also a measure of accuracy of the resulting NE. This measure is normalized to be $0$ if the number of nodes  contributing is $U_*$, and is $1$ if the contributors are $U^*$: more precisely if the number of contributors is $n$ the accuracy is $\frac{n-U_*}{U^* - U_*}$.
Note that we know the value of $U_*$ and $U^*$ for each of the networks that we analyze only because we are in a completely controlled environment.
Finally, we report how all the NE are distributed in the 50 networks, by number of contributors, and how many of them are local minima of the Markov process but not MNE.

\begin{figure}[h]
\begin{center}
\includegraphics*[width=13cm]{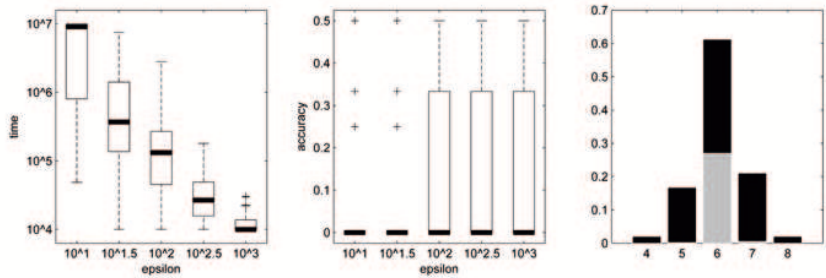}
\caption{Results of the simulated annealing on $50$ random regular networks of $20$ nodes and degree $4$, for $\epsilon$ ranging from $10^1$ to $10^3$.
In the left panel we have the box--plot of the time needed for convergence;
in the center we have the box--plot of the accuracy of the  the algorithm (normalized to be optimal at $0$);
in the right panel we have the distribution of NE (black) in the $50$ networks and those NE (in grey) which are local minima of the Markov process, but not MNE.}   \label{reg_rand_sim_ann}
\end{center}
\end{figure}

\begin{figure}[h]
\begin{center}
\includegraphics*[width=13cm]{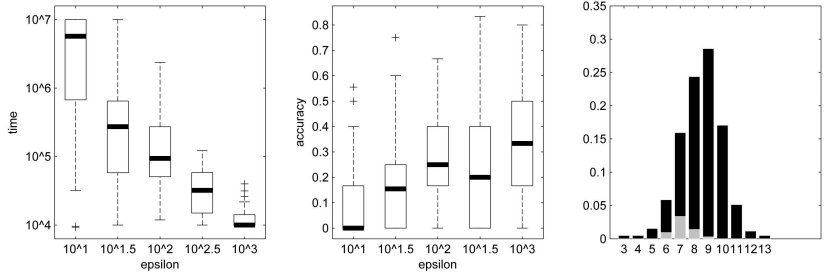}
\caption{Results of the simulated annealing on $50$ random scale free networks of $20$ nodes and average degree $4$, for $\epsilon$ ranging from $10^1$ to $10^3$.
The three plots have the same legend as in Figure \ref{reg_rand_sim_ann}.}   \label{scale_free_sim_ann}
\end{center}
\end{figure}

Results are shown in Figures \ref{reg_rand_sim_ann} and \ref{scale_free_sim_ann}.
The number of time--steps needed for convergence on a single realization of simulated annealing on each of the $50$ different networks are shown as box--plots in the left panels: the thick lines represent the log--median of the realizations,
the edges of the rectangles are first and third log--quartiles, whiskers cover all those observation that would be in the $99\%$ confidence interval (above or below) if the data were log--normal, crosses are outliers outside this range.
The distribution of times is almost the same in the two classes of networks.

The accuracy of simulated annealing is reported in the center panels. 
The scale and the box--plot on the $y$--axis is now linear. 
Simulated annealing performs better on random regular networks, where, even if $\epsilon = 1000 \cdot \bar{\epsilon}$, at least half of the realizations converge to the MNE.
This is clearly not the case for the scale--free networks.
The reason for this is not that the scale--free networks have more local minima.
The right panels report the frequency of all the NE, and that of non--MNE local minima (in grey), as a function of the number of contributors.
For the $50$ random regular networks the variance of contributors between NE is much smaller (the network in Example \ref{reg_rand_ex} is an exception), and even if almost $28\%$ of the NE are non--MNE local minima, the density of contributors they have is very close to the minimal one.
For the $50$ scale--free networks, NE are much more heterogeneous in the number of contributors, and even if only $6\%$ of them are non--MNE local minima, they can have many more contributors than the MNE have.

The main insight from the simulations is that, for some networks, the simulated annealing approach, that we implement in our mechanism, works well and fast even for $\epsilon \gg \bar{\epsilon}$, while this is not the case for others. And this distinction is not trivial.
In regular random networks it is actually very difficult to argue ex--ante who are the contributors in the MNE, because of the full homogeneity between them.
However, a fast version of the mechanism is reasonably accurate in such networks because the Markov process induced by the mechanism itself does not get trapped in the local minima far from the MNE.
On the other hand, in scale--free networks one could approach a NE with a low number of contributors by targeting the hubs, i.e. those nodes with many links.
This strategy will probably find only a local minimum, and this problem arises even when we run our mechanism.
A fast version of the mechanism is not accurate in such networks because such local minima may have many more contributors than the MNE of the network.

\section{Short considerations} \label{conclusion}

The problem of finding a MNE among all the NE is in general not a trivial one, and the difference between the aggregate number of nodes playing $1$ in NE could vary significantly even in homogeneous networks, as shown in Example \ref{reg_rand_ex}.
The star structure (Example \ref{star}) is a trivial but dramatic example: there are two NE, one in which the center alone plays $1$, and another in which all the spokes do so and the center free rides.

The main practical problem in the implementation of the mechanism we propose is clearly the necessity of infinite time.
This paper is only theoretical.
However, simulated annealing is used in practice in many optimization problems.\footnote{%
Crama and Schyns (2003) is a good example related to finance.}
For any $\epsilon>0$ the system will reach a local minimum, which can be easily identified even in finite time (the higher the $\epsilon$ the faster the convergence).
Noting that the values $\epsilon < \bar{\epsilon}$ are typically irrealistically low, and that the algorithm therefor converges very slowly, the choice of a proper heuristic $\epsilon>\bar{\epsilon}$ could be appropriate.
This choice would depend on a profit/costs comparison but also, in the case of finite time, on the structure of the network (e.g. the star needs a single flip to move from the bad NE to the MNE).
As shown in Section \ref{speed}, a particular care should be applied because for some networks there is a concrete risk of finding a local minimum that is very inefficient

Finally, even if the planner does initially not know the real structure of the network, she could infer it link by link as the steps of the mechanism are played.
In this way she could mix the mechanism with a theoretical investigation, and could target nodes non--randomly in order to maximize the likelihood of finding the desired MNE.
The analysis of such a sophisticated approach would be much more complicated.
What we give here is an upper bound that, we prove, holds exactly (even if in the limit of infinite time).
Any improvement on this na\"ive mechanism will work as well, faster, but not in finite \emph{short} time for any possible network, because the original problem is NP--hard.

\setcounter{section}{1} \appendix
\section*{Appendices}
\renewcommand{\thesection}{\Alph{section}}
\renewcommand{\thesubsection}{\Alph{section}.\arabic{subsection}}

\section{Theorem B in Geman and Geman (1984)} \label{Geman}

Geman and Geman (1984) is a pioneering theoretical paper on computer graphics, studying the best achievable quality of images.
Sections X to XII are devoted to the general case of optimization among a finite number of states.
We find there a general theorem (Theorem B at page 731) proving a conjecture on the \emph{Simulated Annealing} heuristic algorithm proposed by Kirkpatrick, Gelatt and Vecchi (1983).
The arising popularity of Simulated Annealing has attested the success of Geman and Geman (1984), which is now cited (according to \emph{scholar.google.com} in January 2010) by almost 10000 papers from all disciplines.

In this appendix we summarize what is necessary for us from this result, with some of the original notation but avoiding most of the thermodynamics \emph{jargon}.
Suppose that there is a finite set $\Omega$ of states, and a function $U:\Omega \rightarrow R_+$, so that, for any $\omega \in \Omega$, $U(\omega)$ is a positive number.
Call $U^* \equiv \max_{\omega \in \Omega} U (\omega)$ the maximal value of $U$, $U_* \equiv \min_{\omega \in \Omega} U (\omega)$ its minimal value, and $\Omega_0 \equiv \arg \min_{\omega \in \Omega} U (\omega)$ those states whose value is $U_*$.
Suppose moreover that we have a fixed transition matrix $X$ between all the elements of $\Omega$ and that this stochastic matrix $X$ is ergodic, i.e. there is a positive probability of reaching any state $\omega' \in \Omega$ from any other state $\omega'' \in \Omega$.
Given any $\omega \in \Omega$, call $X(\omega)$ all those states that can be reached from $\omega$  with positive probability, through $X$, with a single step.

Consider now a discrete time flow with $t=1,2,\dots$ and the following new stochastic process.
$\omega_1$ is any member of $\Omega$.
Imagine that, at time $t$, the process is in the state $\omega_t$, then apply $X$ from $\omega_t$, obtaining a state that we call $\omega_t^{new}$.
We now define $\omega_{t+1}$ as
\begin{equation}
\omega_{t+1} \equiv \left\{
\begin{array}{lll}
\omega_t^{new} & \mbox{with probability} &
\left\{
\begin{array}{ll}
1 & \mbox{if } U(\omega_t^{new}) < U(\omega_t) \ , \\
t^{- \epsilon \left( U(\omega_t^{new})- U(\omega_t) \right) } & \mbox{otherwise;}
\end{array}
\right. \\
\omega_t & \mbox{otherwise.}
\end{array}
\right.
\label{Geman_montecarlo}
\end{equation}
The probability $t^{- \epsilon \left( U(\omega_t^{new})- U(\omega_t) \right) }$ in (\ref{Geman_montecarlo}) identifies the level of acceptance of non--improving changes, which is declining in time at a rate that depends on the constant $\epsilon >0$.
Any such stochastic process will be identified by $\omega_0$ and $\epsilon$: call it $P_{\omega_0, \epsilon}$.

\bigskip

It is easy to prove that at the limit $t \rightarrow \infty$ any realization of $P_{\omega_0, \epsilon}$ will end up in a set of local minima $\Omega_{\epsilon} \subseteq \Omega$.
$\Omega_{\epsilon}$ is such that, for any $\omega', \omega'' \in \Omega_{\epsilon}$ and $\omega_X \in X(\omega')$, $U(\omega')=U(\omega'')$ and $U(\omega') \leq U(\omega_X)$.

\bigskip

The theorem imposes a single condition on $\epsilon$ so that the local minima obtained through $P_{\omega_0, \epsilon}$ are also global minima.

\bigskip

\nn{\bf Theorem B:}
{\it
call $N_{\Omega}$ the cardinality of $\Omega$ and $\Delta \equiv U^*-U_*$.
If $\epsilon<\bar{\epsilon} \equiv \frac{1}{N_{\Omega} \Delta}$, then $\Omega_{\epsilon} \subseteq \Omega_0$ for any realization of $P_{\omega_0, \epsilon}$, independently of $\omega_0$.
}

\bigskip

The proof is by no means trivial, it takes various pages and it is heavily based on the ergodicity of the system.
In Geman and Geman's notation, what they call \emph{temperature} is $\frac{1}{\epsilon \log t}$.
They prove, moreover, that, in the presence of more global minima, the probabilities of ending in any one of them are uniform.

\section{Proof of Lemmas} \label{Proofs}

Consider a finite network and call $x_i \in \{ 0,1 \}$ the action of node $i$, so that $\vec{x}$ is the vector of the actions of all the nodes.
Call $N^1_i$ the set of nodes which are first neighbors of node $i$, and $N^2_i$ those which are second neighbors of node $i$.

We also need the following definitions.
A set of nodes in a network is an \emph{independent set} if, for every link of the network, not both its nodes are in the set.
A set $C$ of nodes in a network is a \emph{covering} if, for every node $i$, $C \cap \left( \{i\} \cup N^1_i \right) \ne \emptyset$ (i.e. if for any node $i$ we consider the set made of $i$ itself and its first neighbors, then at least one of them is also a member of $C$).
A set of nodes in a network is a \emph{maximal independent set} if it is both an independent set and a covering.
In our notation a maximal independent set is characterized by those nodes playing $1$ in a NE $\vec{x}$.

Finally, remember that we have defined the basic transition step $B$ of best response as a Markov process across all states $\vec{x}$, where an unsatisfied node (if existing) is picked with uniform probabilities, and her action is flipped. If there are no unsatisfied nodes $\vec{x}$ is a NE and an absorbing state for $B$.

\bigskip

{\bf Proof of Lemmas \ref{allNE} and \ref{localNE}:}
suppose that $x_i=1$ and we flip her action so that  $x_i^{new}=0$.
Consider now any node $j$ in $N^1_i$, it is clear that $x_j=0$ since $x_i=1$.
All and only new unsatisfied nodes will be all those $j \in N^1_i$ such that $x_k=0$ for any $k \in N^1_j \backslash \{ i \}$.\footnote{%
If this set is empty, then the only unsatisfied node is $i$, but as we will prove below it cannot be the case if we start by applying $F$ to a node who was originally playing $0$.}
If we apply the transition step $B$ to one of them, call her $j$, she will be satisfied again and all her neighbors will be, because if $j$ is such that $x_j=0$ and $x_j^{new}=1$, it is surely the case that any $k \in N^1_j \backslash \{ i \}$ was playing $x_k=0$ and remains at $x_k^{new}=0$.

It could be the case that two such $j$'s that are both neighbors of $i$ and together, are unsatisfied after $i$'s shift.
The fact that one of the two may be chosen instead of the other in an iteration of $B$ is the only random part in the best response rule.

As the neighbors of $i$ are finite $B$ needs to be iterated at most $|N^1_i|$ times and the propagation of best response is limited to $N^1_i$.

\smallskip

{\bf Note:} \emph{a best response from $0$ to $1$ applies only to nodes that are playing $0$, are linked to a node which is shifting from $1$ to $0$, and that node is the only neighbor they have who is originally playing $1$.}

\smallskip

Suppose now that $x_i=0$ is chosen by the stochastic transition $F$, and we flip her action so that $x_i^{new}=1$.
The nodes $j$ in $N^1_i$ who were playing $x_j=0$ will continue to do so, as they will remain satisfied and $B$ will not apply to them.
Any node $j$ in $N^1_i$ (at least one) who was playing $x_j=1$ may be selected by $B$ and will then move to $x^{new}_j=0$.
Note that there is no indeterminacy in how they will be selected by $B$, as they cannot be neighbors together, as they were all playing $1$ in a NE.

By the previous point this will create a propagation, through $B$ to some $k \in N^1_j$, but not $i$, who is now satisfied, and not even to any other $k \in N^1_j \cap N^1_i$, for the same reason.
This proves that the propagation of the best response $B$ is limited to $N^2_i$, and that it ends in a new NE in a number of steps that is at most $|N^1_i \cup N^2_i|$. \fine

\bigskip

{\bf Proof of Lemma \ref{anyNE}:}
we proceed by defining intermediate NE $\vec{x}^1$, $\vec{x}^2$\dots \\
between any two NE $\vec{x}$ and $\vec{x}'$.
$\vec{x}^{n+1}$ will be obtained from $\vec{x}^n$ by flipping one node from $0$ to $1$ (through $F$) and waiting for the best response (the iteration of $B$, which has been proved above to be finite).

\smallskip

If two NE $\vec{x}$ and $\vec{x}'$ are different, it must be that there is at least one $i_1$ such that $x_{i_1}=0$ and $x'_{i_1}=1$ (it is easy to check that any strict subset of a maximal independent set is not a covering any more).
Change the action of that node so that $x_{i_1}^{1}=x'_{i_1}=1$.
By previous proof this will propagate deterministically to $N^1_{i_1}$ and, for all $j \in N^1_{i_1}$, we will have $x^{1}_j = x'_j = 0$.
Propagation may also affect $N^2_{i_1}$ but this is of no importance for our purposes.

\smallskip

If still $\vec{x}^{1} \ne \vec{x}'$, then take another node $i_2$ such that $x^1_{i_2}=0$ and $x'_{i_2}=1$ (${i_2}$ is clearly not a member of $N^1_{i_1} \cup \{ i_1 \}$).
Pose $x_{i_2}^{2}=x'_{i_2}=1$, this will change some other nodes by best response, but not $j \in N^1_{i_1} \cup \{ i_1 \}$, because any
$j \in N^1_{i_1}$ can rely on $x^{1}_{i_1}=1$, and then also $x^{2}_{i_1}=x^{1}_{i_1}=1$ is fixed.

\smallskip

We can go on as long as $\vec{x}^{n} \ne \vec{x}'$, taking any node $i_{n+1}$ for which $x^n_{i_{n+1}}=0$ and $x'_{i_{n+1}}=1$.
This process will converge to $\vec{x}^{n} \rightarrow \vec{x}'$ in a finite number of steps because:
\begin{itemize}
\item when $i_{n+1}$ shifts from $0$ to $1$, the nodes $j \in \bigcup_{h=1}^n \left( N^1_{i_h} \cup \{ i_h \} \right)$ will not change, since they are either $0$--players with a $1$--player beside already (the $1$--player is some $i_h$, with $h \leq n$), or a $1$ (some $i_h$) surrounded by frozen $0$'s;
\item by construction it is never the case that $i_{n+1} \in \bigcup_{h=1}^n \left( N^1_{i_h} \cup \{ i_h \} \right)$, because for all $j \in \bigcup_{h=1}^n \left( N^1_{i_h} \cup \{ i_h \} \right)$ we have that $x^n_{j} = x'_j$;
\item the network is finite. \fine
\end{itemize}

\bigskip

In the above proof, the shift from $\vec{x}$ to $\vec{x}'$ is done by construction re--defining the covering of any $\vec{x}^n$ from the covering of $\vec{x}'$.
It is always certain that, by best response, any $\vec{x}^n$ is also an independent set.

\end{document}